\documentstyle[12pt,aaspp4]{article}

\newcommand {\etal} {{\it et al.}~}

\lefthead{R.M. Sharples {\it et al.}}
\righthead{Spectroscopy of Globular Clusters in NGC~4472}
 
\begin{document}

\title{Spectroscopy of Globular Clusters in NGC~4472}

\author{R.M. Sharples}
\affil{Department of Physics, University of Durham,
Durham DH1 3LE, England.}
\author{S.E. Zepf\altaffilmark{1,2,3}}
\affil{Astronomy Dept, Univ. of California, Berkeley, CA, USA.}
\author{T.J. Bridges}
\affil{Royal Greenwich Observatory, Cambridge, England.}
\author{D.A. Hanes\altaffilmark{2}}
\affil{Physics Dept, Queen's University, Kingston, Ontario, Canada.}
\author{D. Carter}
\affil{Astrophysics Research Institute, Liverpool John Moores University,
Byrom Street, Liverpool, England.}
\author{K.M. Ashman}
\affil{Dept. of Physics \& Astronomy, University of Kansas,
Lawrence, KS, USA.}
\and 
\author{D. Geisler}
\affil{Kitt Peak National Observatory, National Optical Astronomy
Observatories, Tucson, AZ, USA.}
\altaffiltext{1}{Hubble Fellow}
\altaffiltext{2}{Visiting Astronomer, William Herschel Telescope.
The WHT is operated on the island of La Palma by
the Isaac Newton Group in the Spanish Observatorio del 
Roque de los Muchachos of the Instituto de Astrofisica de Canarias.}
\altaffiltext{3}{present address: Astronomy Dept, Yale University, New
Haven, CT, USA.}

\begin{abstract}

Optical multi-slit spectra have been obtained 
for 47 globular clusters surrounding the brightest Virgo elliptical
NGC~4472 (M49). 
Including data from the literature, we analyze 
velocities for a total sample of 57 clusters
and present the first tentative evidence for
kinematic differences between the red and blue cluster populations 
which make up the bimodal colour distribution of this galaxy. 
The redder clusters are more
centrally concentrated and have a velocity dispersion of 240 kms$^{-1}$
compared with 320 kms$^{-1}$ for the blue clusters.
The origin of this difference appears to be a larger
component of systematic rotation in the blue cluster system.
The larger rotation in the more extended blue cluster system 
is indicative of efficient angular momentum transport, as
provided by galaxy mergers.
Masses estimated from the globular cluster velocities
are consistent with the mass distribution estimated from 
X-ray data, and indicate that the M/L$_B$ rises to 
$50$~M/L$_\odot$ at 2.5 R$_e$.

\end{abstract}


\section{Introduction}

The study of extragalactic globular cluster systems can provide
important clues to the formation history of their host galaxies.  This is
particularly true for elliptical galaxies for which there are two
currently popular paradigms.  
One paradigm is the standard
monolithic collapse model in which elliptical galaxies form in a
single burst of star formation at high redshift (e.g. Arimoto \&
Yohii 1987). In contrast, hierarchical structure formation
theories predict that spheroidal galaxies form continuously through a
sequence of galaxy mergers (e.g. Cole \etal 1994; Kauffmann 1996).  Ashman
\& Zepf (1992) explored the properties of globular clusters in models in
which elliptical galaxies are the products of the mergers of spiral
galaxies, and showed that the greater specific frequency of globular
clusters around ellipticals relative to spirals could be explained if
globular clusters form during the mergers. They also predicted
that elliptical galaxies formed by mergers will have two or more
populations of globular clusters - a metal-poor population associated with
the progenitor spirals, and a metal-rich population formed during the
merger.  In contrast, monolithic collapse models naturally produce
unimodal metallicity distributions.  The discovery that the
globular cluster systems of several elliptical galaxies have bimodal
colour (and by implication metallicity) distributions (Zepf \& Ashman
1993; Whitmore \etal 1995;  Geisler \etal 1996) provides strong support
for the merger model. Geisler \etal (1996)  
and Lee et al. (1998) also show that the red
(metal-rich)  cluster population is more centrally concentrated than the
blue (metal-poor) population, as predicted by Ashman \& Zepf (1992).
 
Recently, an alternative view has been presented by Forbes
et al.\ (1997), who suggest that the bimodal color distributions
may not be due to mergers, but to a multi-phase single collapse.
Although the primary physical mechanism known to produce distinct 
formation episodes is mergers, it is important to attempt
to distinguish between these competing models for the formation of
globular cluster systems and their host elliptical galaxies.
The kinematics of globular cluster systems may offer such a test
of these models. In the multi-phase collapse picture, angular momentum
conservation requires that the spatially concentrated metal-rich
population rotates more rapidly than the extended metal-poor
population. In contrast, simulations of merger models indicate that
mergers
typically provide an efficient means of angular momentum tranfer,
and that the central regions have specific angular momentum that
is lower than the outer regions (Hernquist 1993; Heyl, Hernquist and 
Spergel 1996).

Studies of the kinematics of globular cluster systems therefore
provide important constraints on the formation history of elliptical
galaxies. They also provide useful probes of the mass distribution
of elliptical galaxies at radii larger than can be reached by
studies of the integrated light. The extended nature of globular 
cluster systems allows the dynamical mass determined from their velocities
to be compared at similar radii to masses determined through studies
of the hot X-ray gas. A recent example is the
study of the M87 globular cluster system by Cohen \& Ryzhov (1997),
who find that a rising mass-to-light ratio is required out to radii
of $\sim 3 R_e$, in agreement with X-ray mass determinations. 
However, M87 occupies a privileged position at the center of the 
Virgo cluster, so it is critical to test whether the rising
mass-to-light ratio, and the agreement with X-ray masses, is true
for more typical cluster elliptical galaxies.

In this paper we present a 
spectroscopic study of the globular cluster system of the
elliptical galaxy NGC~4472 (M49). This is the brightest elliptical
galaxy in the Virgo cluster and has been the subject of a detailed
photometric study by Geisler \etal (1996).
The only previously published spectroscopic data for the
NGC~4472 globular cluster population is by Mould \etal (1990) 
who presented velocities and line strengths for 26 clusters.

The outline of our paper is as follows:
Section 2 discusses the sample selection 
together with our observations
and data reduction; Section 3 discusses the kinematics of
the metal-rich and metal-poor populations in the 
context of the merger model, and
analyses the implications for the overall M/L ratio in NGC~4472. 
Finally, we present our conclusions in Section 4.

\section{Observations and Data Reduction}

\begin{figure}
\plotone{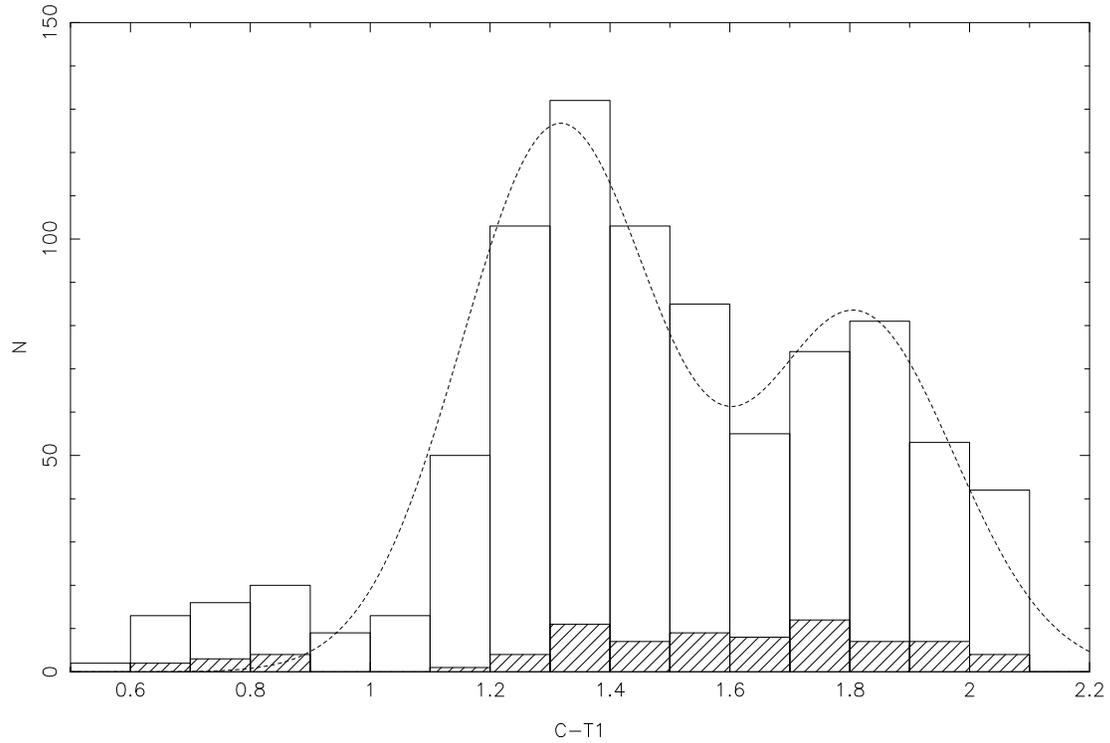}
\caption{Colour distribution for globular cluster
candidates with $19.5<{\rm V}<22.5$ and $0.5<{\rm C-T}_1<2.2$
from Geisler \etal (1996). The open histogram is for
the full sample of 860 candidates from which our spectroscopic targets
were selected; the shaded histogram is for the 79 objects
for which spectra were obtained. \label{fig1}}
\end{figure}

Geisler \etal (1996) 
have made a deep (R$\sim 25$) photometric study of the
globular cluster system surrounding NGC~4472 in the
integrated Washington CT$_1$ system. They showed that the
colour distribution was clearly bimodal and
could be well-fit by two Gaussians with peaks 
at $C-T_1 = 1.32$, and $C-T_1 = 1.81$, 
corresponding to metallicities of [Fe/H]$=-1.3$
and $-0.1$. We selected our sample for spectroscopic 
study from a preliminary
version of the Geisler \etal (1996) catalogue.
In order to avoid
any biases in the spectroscopic cluster sample, particularly
with regard to the presence of any young
cluster population, we have
applied only a very broad colour
cut of $0.5<C-T_1<2.2$ together with a magnitude cut of $19.5\leq {\rm V} 
\leq 22.5$ to the original 
catalogue, where V$\simeq$T$_1+0.5$ (Geisler 1996).
The colour distribution of this sample is shown in 
Fig.~\ref{fig1}.
Because the catalogues are incomplete near the bright
central parts of the galaxy, the multi-slit masks were
offset along the major axis of the galaxy by $\sim 3'$
in order to maximise the spatial and spectral coverage of the samples.

Spectroscopic observations of 79 cluster candidates
were obtained with the Low Dispersion
Survey Spectrograph (Allington-Smith \etal 1994) on
the 4.2m William Herschel Telescope
in April 1994. 
Further details of the observing setup are given
in Table~\ref{tab1} and in Bridges \etal (1997).
Dome and twilight flats were taken
at the beginning and end of each night, and 
the spectra were wavelength calibrated 
using frequent CuAr arcs.
Long-slit spectra of the Galactic
globular clusters M92 (NGC~6341;
[Fe/H]=-2.24),  M13 (NGC~6205; [Fe/H]=-1.65) \&
NGC 6356 ([Fe/H]=-0.54), and the radial velocity standards 
HD194071 (G8III) \& HD132737 (K0III) were taken for velocity
and metallicity calibration. 
Data reduction for the first mask followed closely the procedures detailed
in Bridges \etal (1997) using the LEXT (Allington-Smith \etal
1994) software package.
A 3rd order polynomial fit to the arc spectra gave
residuals $\sim 0.13$\AA (8~kms$^{-1}$) and the spectra were rebinned
to linear and logarithmic wavelength scales over
the wavelength range $3800-6000$\AA ~with a bin size of
$2.0$\AA.  The spectra
were then optimally extracted and sky subtracted, using
linear fits for the background sky.  

For the other
two masks a slightly different procedure was followed.
Rather than adjust the individual multi-slit lengths to fill the available
mask dimensions, a fixed-length multi-slit of $8''$ was
adopted and the targets were nodded up and down the slits
by $\pm 2''.5$ on consecutive exposures. This procedure allowed
us to substantially increase the number of objects observed
per mask and, by differencing pairs of exposures to subtract the
sky background, removed some of the systematic effects due to
flatfielding and irregularities in the slit profiles during this
crucial stage. Comparison of spectra for objects observed
on more than one mask ($\S 2.1$) indicated that this technique does
not degrade the S/N achieved and substantially increased the
efficiency of our observations.


\subsection{Radial Velocities and Confirmed Globular Clusters}

\begin{figure}
\plotone{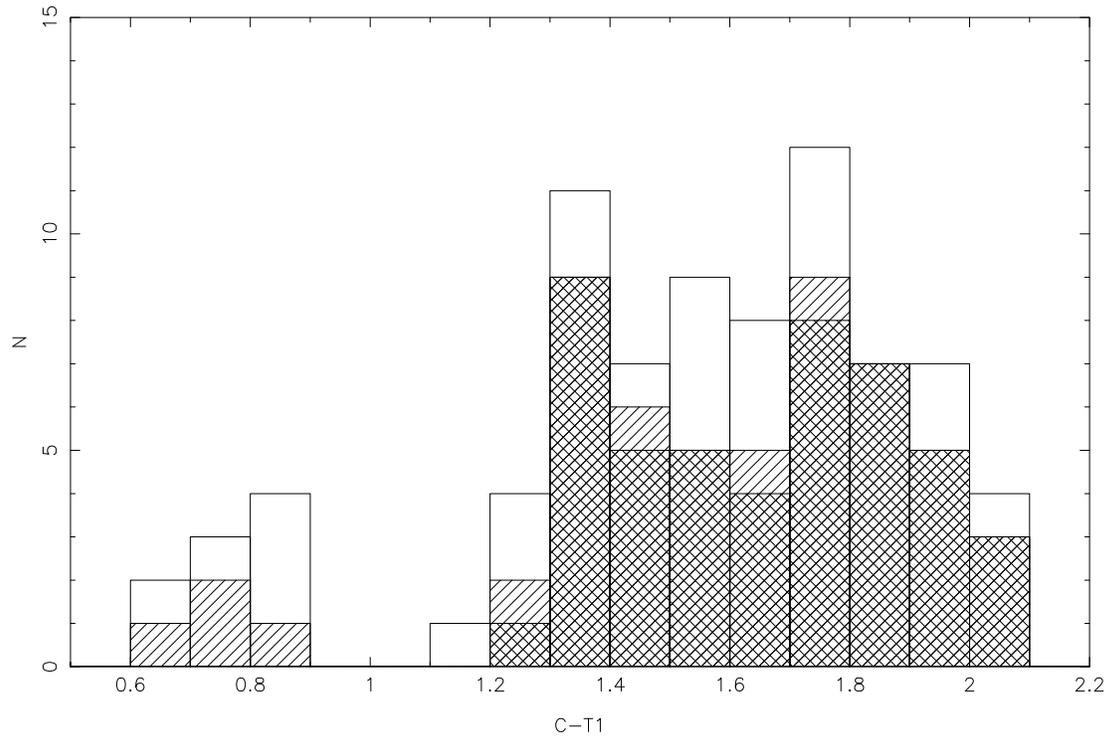}
\caption{The open histogram shows the colour distribution for 
the 79 globular cluster candidates for which spectra were obtained.
The shaded histogram shows the 55 candidates for which reliable
velocities were derived from the cross-correlation analysis. The
cross-hatched area shows the colour distribution of the 47 objects
identified as clusters 
on the basis of their radial velocities in Table~\ref{tab2}.
\label{fig2}}
\end{figure}

Radial velocities were obtained by cross correlating with
the template spectra of M13, M92, NGC 6356, HD132737
and HD194071, and forming a weighted average of the
results after rejecting poor matches (normalised
correlation amplitudes $<0.15$). Velocities for the template clusters
were taken from Hesser \etal (1986).
Cross-correlation of spectra from
the twilight sky frames indicated an rms velocity
uncertainty of 37 kms$^{-1}$ for high S/N data; this
was added in quadrature to the scatter between templates
to give the final velocity errors in Table~\ref{tab2}.
Eleven objects were measured on more than one mask and have an
rms velocity difference of 54 kms$^{-1}$, which
is consistent
with the estimated velocity errors.
Table~\ref{tab2} contains positions, colour and magnitudes
for 79 cluster candidates for which spectra were obtained, together with
velocities and velocity errors for 55 objects with reliable
cross-correlation results (generally against 3 or more templates).
The astrometric solution was obtained from a fit to 43
secondary standards distributed across the field
and the relative positions should be good
to $\pm 0''.3$. The colour distribution of the 55 objects
with radial velocities is shown in Figure~\ref{fig2}.

\begin{figure}
\plotone{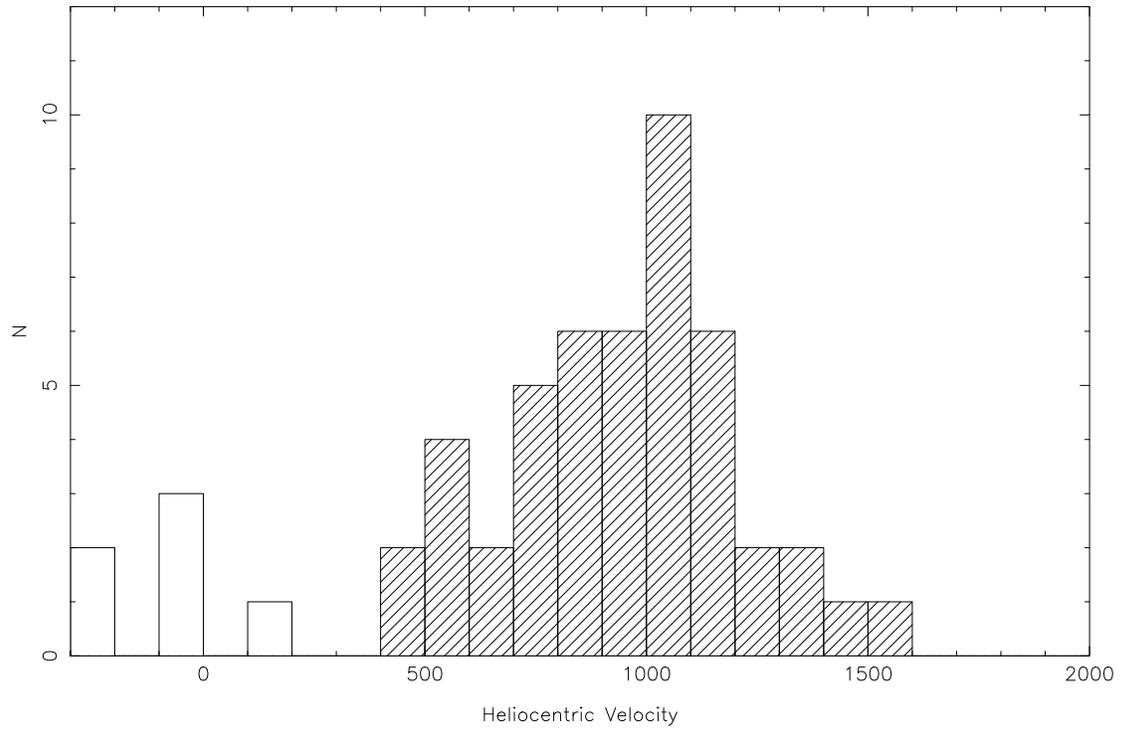}
\caption{Heliocentric velocities for 53 globular cluster candidates
around NGC~4472. Two objects (presumably compact galaxies) with
velocities $>10000$ kms$^{-1}$ have been omitted. The shaded
histogram shows those candidates assumed to be bona fide
clusters belonging to NGC~4472 (see text for details).
\label{fig3}}
\end{figure}

A histogram of the radial 
velocities of these 55 objects
is given in Figure~\ref{fig3}. Following Mould \etal (1990) we take
the velocity range $300<V_{he}<2000$ kms$^{-1}$ as representative
of globular clusters belonging to NGC~4472
which has a recession velocity of 961  kms$^{-1}$
(Sandage \& Tammann 1981); the 47 objects in
this range have a mean velocity of $943\pm38$ kms$^{-1}$ and a
velocity dispersion of 258 kms$^{-1}$. Assuming $\sigma=110$ kms$^{-1}$
for halo subdwarfs (Mould \etal 1990), 
the lowest velocity object in our cluster
sample (2163) has a 4\% chance of belonging to the NGC~4472 system
and only 0.02\% of being a Galactic star; the highest velocity
object excluded (7731) has a 17\% chance of being a Galactic star and
only 0.2\% chance of belonging to the NGC~4472 system.
Including or excluding these objects from the cluster
sample gives a velocity dispersion in the range  $248-279$ kms$^{-1}$.

  
\section{Results}

The only previous spectroscopic study of the NGC~4472 globular
cluster system is that of Mould \etal (1990),
who obtained spectra for 54 candidates from which
they identified 26 clusters. The contamination by foreground
stars and background stars in their study (49\%) is higher
than that obtained here (15\%), presumably due to the magnitude 
and colour range adopted. Thirteen objects in Table~\ref{tab2}
are in common with the Mould \etal (1990) sample; for twelve of these
the radial velocity measurements are in good agreement, with
a mean difference of $-2$ kms$^{-1}$ and a dispersion
of 79 kms$^{-1}$. The remaining object (5090) has a velocity
of 582 kms$^{-1}$ in Table~\ref{tab2} and 959 kms$^{-1}$ in
Mould \etal (1990); although we only have one spectrum of this
object, there are good cross-correlations with all of the
templates and it seems likely that the source of the discrepancy
is a mis-identification. 

Table~\ref{tab3} contains data for
the combined sample of 57 clusters. The mean velocity error
for the overlap sample from our measurements is 46 kms$^{-1}$,
so we have adopted a typical $1\sigma$ error of $\sqrt (79^2-46^2)=65$
kms$^{-1}$ for those clusters with velocities from Mould \etal (1990).
Where clusters have velocities available from our LDSS-2 data, we list
these in preference because of their smaller formal errors; 
however, the results presented below do not change significantly 
if instead the two datasets are averaged.
Also indicated in this table is the angular distance of each
cluster from the centre of NGC~4472 (taken as RA=12~27~15.0
Dec=+08~16~45 (1950)) and the projected distance along the major
axis of the elliptical isophotes of the galaxy (taken as
PA=162$^o$; Sandage \& Tamman 1981). The [Fe/H] values are
taken from Geisler \etal (1996).


\subsection{Kinematic Properties}

One of the most important characteristics of the NGC~4472
globular cluster system is the clear bimodal nature of its
colour (metallicity) distribution. Geisler \etal (1996)
have already demonstrated structural differences between
the metal-rich and metal-poor  populations, with the
metal-rich clusters being significantly more concentrated
to the centre of NGC~4472 than their metal-poor counterparts.
The primary goal of our
study was therefore to search for any kinematic differences
between the metal-rich and metal-poor globular cluster populations.

\begin{figure}
\epsscale{0.8}
\plotone{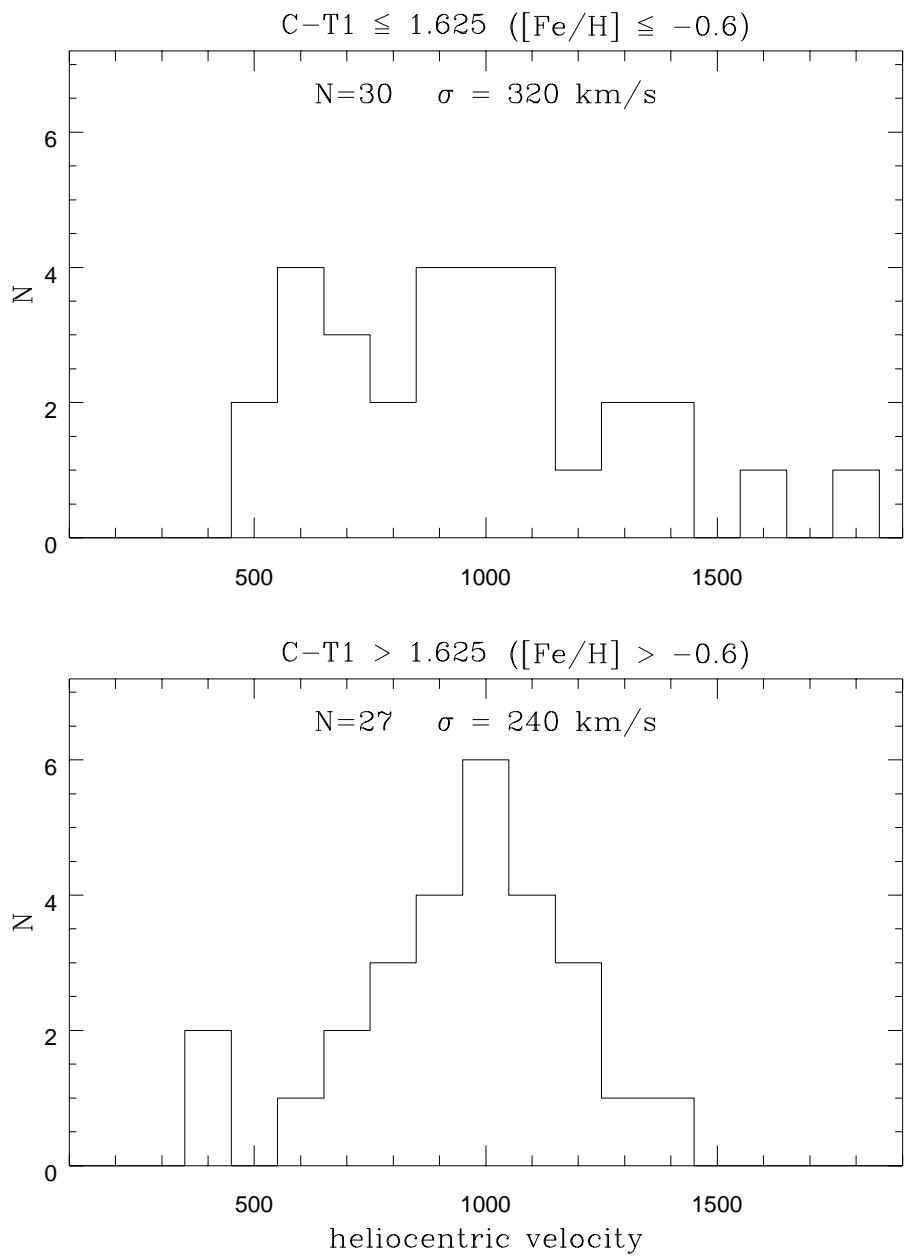}
\caption{A comparison of the velocity distributions
of the metal-poor and metal-rich cluster populations in 
NGC~4472.
An F-test rejects the hypothesis that these two have the same
dispersion at the $86\%$ confidence level. The mean velocities
of the two populations $969\pm 58$ kms$^{-1}$ (N=30) and
$946\pm 46$ kms$^{-1}$ (N=27) are identical to within the errors.
\label{fig4}}
\end{figure}

In order to divide the sample, we have run a KMM mixture-modelling
test (Ashman \etal 1994) which indicates that the
most probable boundary between the 
metal-rich and metal-poor clusters is at
C-T$_1$=1.625 or [Fe/H]=-0.57  (c.f. Fig.~\ref{fig1}).
Fig.~\ref{fig4} shows the velocity histograms for the 57 clusters
in Table~\ref{tab3} divided in this way. The two distributions do
indeed appear to be different with the
metal-rich clusters having a lower velocity
dispersion; an F-test rejects the hypothesis that the 
two populations have the same
dispersion at the $86\%$ confidence level.
Although only tentative given the current sample size, this is a potentially
powerful result, and would  
represent the first detection of a velocity dispersion
difference in a bimodal globular cluster system. In particular it
reinforces the claim that the metal-rich and
metal-poor populations are distinct
(e.g.\ Zepf \& Ashman 1993, Geisler et al.\ 1996).
We investigate the possible origins of these differences further below.

\subsubsection{Rotation Velocity}

\begin{figure}
\plotone{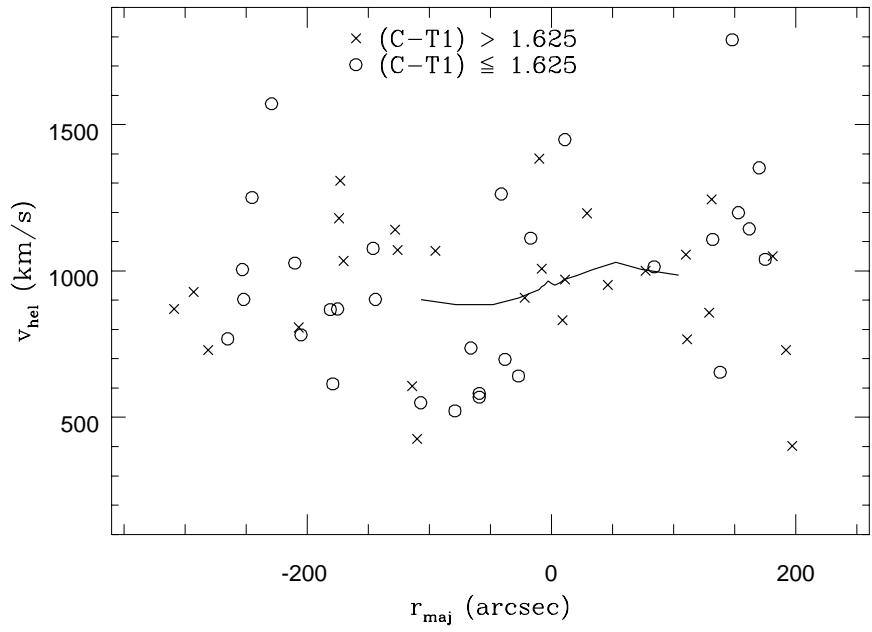}
\caption{A plot of the velocities of globular
clusters against their distance projected along the major axis
of NGC~4472. The solid line shows the rotation
of the stellar component of the galaxy from Fisher \etal (1995)
normalized to the systemic velocity of the galaxy 
from Sandage \& Tamman (1981).
\label{fig5}}
\end{figure}

One of the key observations for testing galaxy formation models is
the amount of systematic rotation in globular cluster subsystems. Since
mergers are believed to be effective in transporting angular momentum
outwards, this provides a mechanism by
which the resulting cluster population can form
a dynamical system primarily
supported by anisotropy rather than rotation.
This natural route to hot, slowly rotating
dynamical systems (at least near the centre) 
does not arise in a (modified) monolithic collapse picture
where dissipation is included to account
for the concentrated distribution of metal-rich clusters compared
to metal-poor clusters.
Figure~\ref{fig5} shows the velocities of the globular
clusters in Table~\ref{tab3} plotted against their 
radial distance projected along the major axis of NGC~4472. 
Formally, the least squares fits to
the metal-poor and metal-rich subpopulations are $0.57\pm0.39$ 
kms$^{-1}$/arcsec (Spearman rank correlation coefficient 0.26)
and $0.10\pm0.31$ kms$^{-1}$/arcsec (correlation 
coefficient -0.07). However, this method assumes that the line of
nodes of the cluster system is the same as that of the galaxy, which may
not be the case if the system has undergone a major merger.

To constrain the rotation amplitude of
the whole cluster system more generally, we have fitted a function
of the form 
\begin{equation}
V(r)=V_{rot}\sin (\theta-\theta_0)+V_o
\end{equation}
to the radial velocity and position angle data using a
non-linear least-squares algorithm. The resulting amplitude
of rotation $V_{rot}$ is $85\pm 35$ kms$^{-1}$
at position angle $\theta_0=125\deg \pm 30\deg$, in the same sense
as the rotation of the stellar component (NE approaching, SE
receeding).
This is consistent with, but somewhat smaller than the value
of 113 kms$^{-1}$ found by Mould \etal (1990)
using a smaller sample. To check the reality of this result
we have repeated the fit after randomizing the position angles
of the clusters; in 50 trials, 12\% of the fits had
a rotation amplitude greater than or equal to $85$ kms$^{-1}$,
giving a confidence level of 88\% that the detected
rotation is real.
Since the significance of this result is not very large, we have
attempted to put an upper bound on the rotation of the whole
cluster system using Monte-Carlo simulations. By generating
artificial samples with the same positional angle 
distribution as the data, but with velocities
selected from a fixed velocity amplitude 
chosen to lie in the range from $50-250$
kms$^{-1}$ and a fixed dispersion of 258 kms$^{-1}$, we can put
an upper limit on $V_{rot}$ of 150  kms$^{-1}$ at the 
95\% confidence level. As noted by Mould \etal (1990), one
of their clusters (\# 19) has a significantly larger
radial velocity than any of the other clusters and should probably
be excluded from the fit; Table~\ref{tab4} shows the 
corresponding values excluding this cluster. 

\begin{figure}
\epsscale{0.8}
\plotone{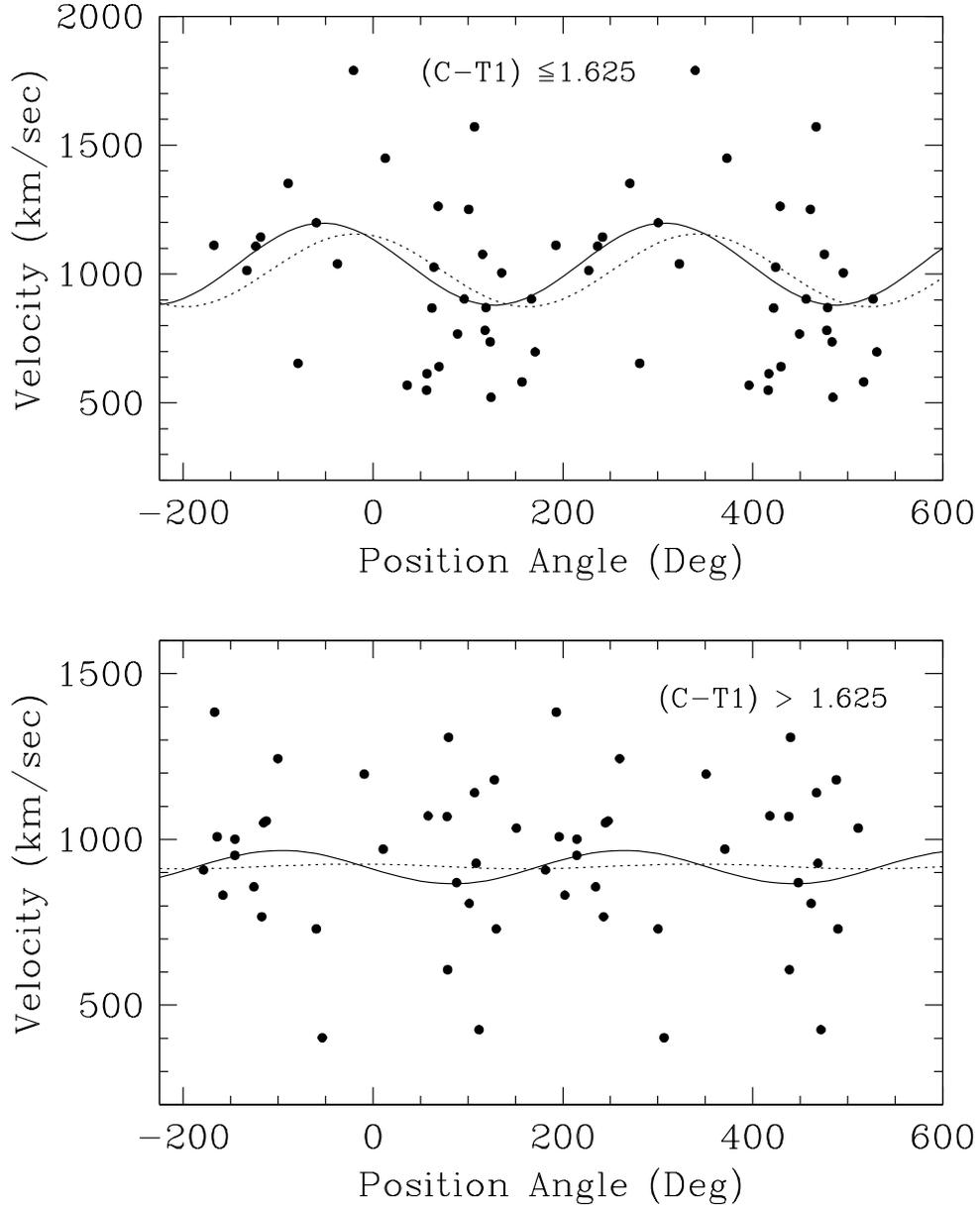}
\caption{A plot of the velocities of globular
clusters against their position angle (measured N through E).
The major axis of NGC~4472 is at PA$=162^o$ and two complete phases
are shown for clarity. The upper panel shows the
results for the metal-poor (C-T$_1\leq 1.625$) clusters; the
lower panel shows the metal-rich (C-T$_1> 1.625$) clusters.
Non-linear least squares fits
of equation 1 are shown by the solid lines;
the dashed lines show the best fit if the
position angle is constrained to be $162^o$.
\label{fig6}}
\end{figure}

Figure~\ref{fig6} shows the results obtained when
these fits are made to the metal-rich and 
metal-poor subsamples separately. Although the sample size is small,
there is evidence that the metal-poor (blue) clusters
rotate faster than the metal-rich (red) clusters. This is
opposite to what is expected in a conventional collapse
picture, where the natural outcome is an old, slowly rotating, metal-poor
halo population of clusters, surrounding a younger, 
more concentrated, population of metal-rich clusters
which has spun-up due to conservation of angular momentum
(c.f. the situation in our own Galaxy).
Mergers however provide a natural mechanism for
transfer of angular momentum from the metal-rich to the metal-poor
clusters. There is also a hint in Table~\ref{tab4} that the rotation axis
for the metal-rich clusters may not be the same as that of the
stellar component (PA$=162\deg$) and the blue clusters. This is
also what might be expected in the merger picture, although given
the small number of clusters, the statistical significance is weak.
 
\subsection{Velocity Dispersion}

\begin{figure}
\plotone{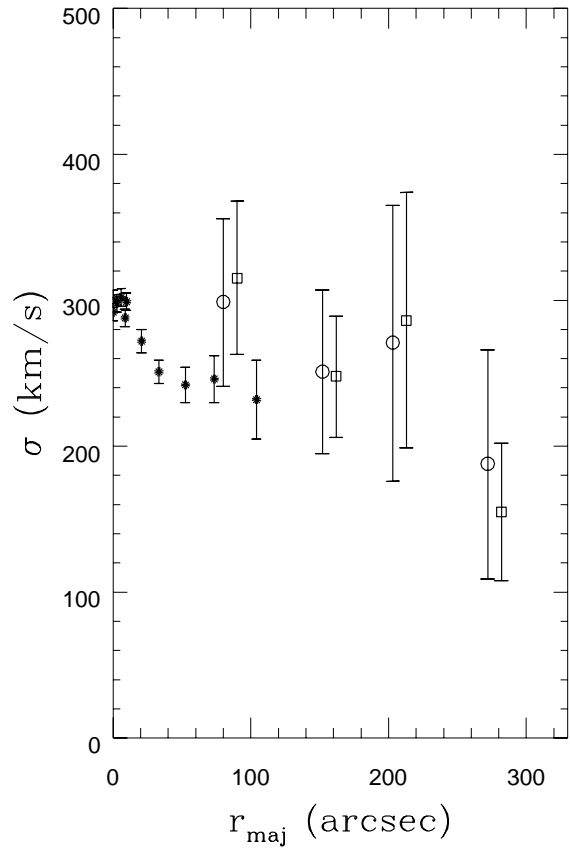}
\caption{A plot of the velocity dispersion of the globular
clusters in NGC~4472 against the projected distance along the major
axis. Open circles are for the full sample, whilst
open squares are excluding cluster \# 19. The velocity
dispersion of the stellar light is shown by the solid symbols.
The importance of globular cluster samples in probing the 
outer haloes of ellipticals is clearly shown in this plot.
\label{fig7}}
\end{figure}

Figure~\ref{fig7} illustrates the velocity dispersion
profile of the globular cluster system of NGC~4472. 
Flat rotation curves with the amplitudes given in Table~\ref{tab4}
have been subtracted from the blue and red populations
independently. The open circles show the robust estimators
of velocity dispersion and errors using the ROSTAT code (Bird \&
Beers 1993) with all the data included; the open squares
show the effects of removing cluster \#19 from the sample (note
that although cluster \#19 is only in the final radial bin, 
removing it from the sample affects all the points, since the
rotation velocity correction is also different c.f. Table~\ref{tab4}).
Also plotted are the stellar velocity dispersion data for
the integrated light of the galaxy from Fisher \etal (1995).
The clusters appear to form a hotter dynamical population than
the halo stars, but unlike the case of M87 (Cohen \& Ryzhov 1997),
the dispersion profile does not increase with radius. Because the
clusters and stars are known to have different structural profiles
(Lee \etal 1998), they may still, of course, be in dynamical
equilibrium with a single halo mass distribution.

\subsection{Mass-to-Light Ratio}

\begin{figure}
\plotone{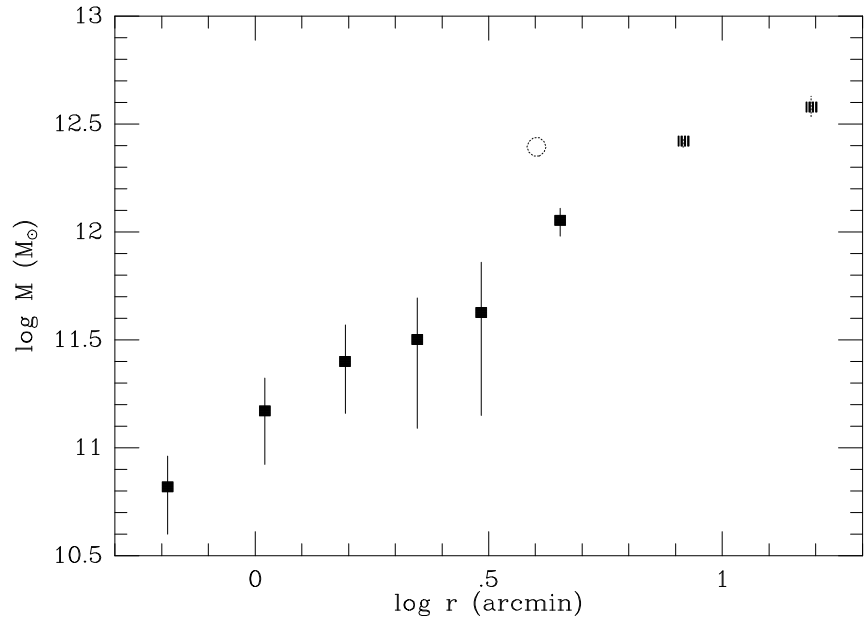}
\caption{Integrated mass distribution for NGC~4472. The solid 
symbols (with error bars) are
the X-ray estimates from Irwin \& Sarazin (1986). The open circle
shows the projected mass estimator applied to the globular cluster
population.
All mass estimates have been scaled to a Virgo
cluster distance of 16~Mpc.
\label{fig8}}
\end{figure}

The halo mass distribution of elliptical galaxies is
poorly known because of the lack of easily observed tracers
such as cold HI gas. Globular clusters provide a potentially
very important probe of the outer regions of ellipticals (e.g. Cohen \&
Ryzhov 1997) to complement the use of planetary nebulae
(Ciardullo \etal 1993) and X-ray gas (Nulson \& B\"{o}hringer 1995).
Fig~\ref{fig8} shows the integrated mass distribution 
for NGC~4472. The solid points
are X-ray estimates from Irwin \& Sarazin (1996). Open circles
show the projected mass estimator of 
Heisler \etal (1985) applied to our globular cluster sample
assuming isotropic orbits and an extended mass distribution. 
The assumption of a point mass distribution would
decrease the projected mass estimator by a factor $\sim 2$, but
this assumption is clearly inconsistent with the 
extended mass profile implied by the X-ray data. More likely, the
slightly higher mass estimate from the globular cluster data
reflects a tangential anisotropy in the orbits, and there
is some evidence to support this in our rotation analysis of
the clusters. Correcting the velocities for our best-fit
rotation solution decreases the projected mass estimator by $\sim 20\%$. 
More importantly, any tendency for the cluster populations to
have {\em radially} anisotropic orbits, as might be expected in a 
monolithic collapse scenario, would increase the 
projected mass estimator even further 
above that derived from the X-ray gas.

We have calculated the M/L$_B$ ratio using a standard B-band
growth curve from the RC3 (de Vaucouleurs et al 1991),
assuming a distance to Virgo of 16 Mpc. At a projected
distance of $\sim 2.5$ R$_e$ ($4'$), the projected mass estimator 
gives M$ \approx 2.5\; 10^{12}$~M$_\odot$ which implies
M/L$_B \approx 50$~M/L$_\odot$. This may be compared
with a value of  M/L$_B \approx 7$~M/L$_\odot$ derived by 
Saglia \etal (1993) from the stellar kinematics of the
integrated light. Clearly these kinematic studies indicate that the
M/L ratio is increasing rapidly
with radius, and support the conclusion from X-ray measurements
that NGC~4472 has an extended dark matter distribution
similar to (although possibly not as extreme as) that seen in M87.

\section{Conclusions}

We have made a detailed spectroscopic study of the globular
cluster system of NGC~4472, and have more than doubled the number of 
confirmed clusters to 57. Whilst this remains a statistically
small sample, the data show several interesting properties
when combined with the accurate colour/metallicity data
from Geisler \etal (1996). When the complete sample is
divided into a metal-rich (47\%) and a
metal-poor (53\%) subset on the basis of their bimodal
colour histogram, the metal-poor subset appears to have
a broader distribution of velocities. We have investigated this
further, and conclude that the most likely cause is a higher
mean level of rotation in the metal-poor cluster system, which
is consistent with that of the underlying stellar halo in 
amplitude and position angle (but with a much higher specific
angular momentum). The metal-rich clusters on the other hand
show only weak evidence for any rotation, and 
about an axis which is tilted $\sim 50^o$ from that of
the other components. These results are qualitatively in
agreement with the predictions of a model in
which the metal-rich clusters are formed during the
merger of two massive gas-rich galaxies, each with its
own old metal-poor cluster population. The cluster system of NGC~4472
forms a dynamically hotter population than the stellar
halo, but is consistent with being in dynamical equlibrium with
the halo potential defined by the hot X-ray emitting plasma, and
supports the
presence of a dark $\sim 10^{12}{\rm M}_\odot$ halo in 
this giant elliptical galaxy.

\acknowledgements

We thank Ken Freeman for useful discussions,
Karl Glazebrook for his updates to the Unix
version of the LEXT data reduction package, and Jeremy Mould for
providing finding charts to identify previously observed clusters.
SEZ acknowledges receipt of a Hubble Fellowship. 
DAH would like to acknowledge support through an
Operating Grant awarded by the Natural Sciences and
Engineering Research Council of Canada. This work
was undertaken with the support of  NATO 
Collaborative Research Grant No. 941223 and is
supported in part by NASA through grant No. GO-06699.01-95A (to D.G.) 
from the Space Telescope Science
Institute, which is operated by the Association of Universities for
Research in Astronomy, Inc., under NASA contract NAS5-26555.

\begin{deluxetable}{ll}
\tablewidth{0pt}
\tablenum{1}
\tablecaption{Observing Log. \label{tab1}}
\startdata
 Dates & April 11-14, 1994 \nl
 Telescope/Instrument & 4.2m WHT/LDSS-2 \nl
 Dispersion (Resolution) & 2.4 \AA/pixel (6 \AA ~FWHM) \nl
 Detector & 1024$^2$ TEK CCD \nl
 Wavelength Coverage (max) & 3800--6000 \AA \nl
 Seeing & 1--2$^{\prime \prime}$ \nl
 Exposure time (Mask \#1/\#2/\#3) & 3.0/3.5/3.5 hr \nl
 Number of Objects (Mask \#1/\#2/\#3) & 21/38/38 \nl
\enddata
\end{deluxetable}

\clearpage

\begin{deluxetable}{lllrrllclllrrll}
\tablenum{2}
\tablecaption{Velocities of
globular cluster candidates in NGC~4472.  Successive columns
give ID, T1 magnitude, 
C-T1 colour (all from Geisler \etal 1996), heliocentric velocity, 
velocity error, RA(1950) and Dec(1950). \label{tab2}} 
\scriptsize
\startdata
ID  &    T1      &   C-T1   &   V(he)   &  Err   &     RA(1950)     & Dec(1950) & &
ID  &    T1      &   C-T1   &   V(he)   &  Err   &     RA(1950)     & Dec(1950)\nl
\tableline
1407    &  21.89    &  1.28   &     -8    &  44    &   12 27 04.91   &    8 12 46.4 & &5263    &  19.91    &  0.77   &    -88    &  40    &   12 27 03.22   &    8 17 30.0 \nl
1483    &  20.75    &  1.56   &           &        &   12 27 12.94   &    8 12 57.8 & &5323    &  20.33    &  0.82   &           &        &   12 27 16.29   &    8 17 33.8 \nl
1518    &  19.25    &  1.85   &   1050    &  36    &   12 27 07.89   &    8 13 00.5 & &5511    &  21.17    &  1.74   &           &        &   12 27 17.31   &    8 17 44.7 \nl
1650    &  20.85    &  1.95   &           &        &   12 27 23.29   &    8 13 13.4 & &5561    &  20.82    &  1.39   &    903    &  48    &   12 26 56.84   &    8 17 48.3 \nl
1712    &  20.36    &  1.34   &   1144    &  40    &   12 27 07.54   &    8 13 18.9 & &5629    &  21.09    &  1.36   &    522    &  52    &   12 27 11.90   &    8 17 52.2 \nl
2031    &  20.71    &  1.37   &           &        &   12 27 15.12   &    8 13 46.3 & &5673    &  21.73    &  1.25   &           &        &   12 27 20.81   &    8 17 54.2 \nl
2045    &  20.94    &  1.77   &    857    &  54    &   12 27 06.50   &    8 13 47.7 & &5943    &  21.07    &  1.56   &           &        &   12 27 11.46   &    8 18 11.6 \nl
2140    &  20.45    &  1.80   &    730    &  53    &   12 27 21.64   &    8 13 55.3 & &6164    &  19.79    &  1.65   &    426    &  30    &   12 27 12.25   &    8 18 27.1 \nl
2163    &  20.15    &  2.01   &    402    &  43    &   12 27 23.34   &    8 13 57.6 & &6284    &  19.44    &  1.57   &    569    &  54    &   12 27 25.39   &    8 18 36.8 \nl
2178    &  21.51    &  1.19   &           &        &   12 27 04.82   &    8 13 59.0 & &6294    &  21.02    &  1.64   &   1034    &  84    &   12 27 01.30   &    8 18 37.7 \nl
2306    &  20.35    &  1.62   &           &        &   12 27 25.13   &    8 14 09.0 & &6427    &  21.11    &  1.79   &   1141    &  50    &   12 27 12.47   &    8 18 47.5 \nl
2341    &  20.76    &  1.91   &   1001    &  68    &   12 27 00.21   &    8 14 12.9 & &6520    &  20.06    &  1.86   &    607    &  57    &   12 27 16.75   &    8 18 53.5 \nl
2406    &  20.84    &  2.03   &   1244    &  70    &   12 27 13.23   &    8 14 18.4 & &6564    &  20.03    &  1.34   &   1077    &  31    &   12 27 10.77   &    8 18 57.2 \nl
2482    &  21.58    &  2.08   &    767    &  56    &   12 27 10.16   &    8 14 24.6 & &6696    &  20.08    &  1.59   &    550    &  52    &   12 27 21.45   &    8 19 08.5 \nl
2543    &  20.27    &  1.36   &   1199    &  48    &   12 27 20.33   &    8 14 29.3 & &6721    &  21.09    &  1.73   &   1180    &  45    &   12 27 07.36   &    8 19 10.5 \nl
2569    &  20.12    &  1.89   &   1056    &  46    &   12 27 11.32   &    8 14 31.7 & &6872    &  20.15    &  1.46   &    870    &  41    &   12 27 09.21   &    8 19 20.7 \nl
2634    &  19.70    &  1.56   &   1014    &  57    &   12 27 07.08   &    8 14 37.9 & &6989    &  20.61    &  1.75   &   1071    &  50    &   12 27 21.98   &    8 19 30.5 \nl
2679    &  21.38    &  1.59   &           &        &   12 27 19.00   &    8 14 41.5 & &7174    &  21.01    &  1.51   &           &        &   12 27 20.43   &    8 19 46.3 \nl
2757    &  19.62    &  1.45   &    -79    &  37    &   12 27 08.35   &    8 14 47.2 & &7197    &  20.94    &  1.50   &    782    &  50    &   12 27 08.44   &    8 19 48.2 \nl
2860    &  20.27    &  1.21   &           &        &   12 27 22.00   &    8 14 52.5 & &7340    &  20.91    &  1.77   &   1308    & 124    &   12 27 17.39   &    8 19 59.0 \nl
3119    &  20.78    &  1.74   &  11068    &  50    &   12 27 15.44   &    8 15 09.9 & &7399    &  20.35    &  1.40   &   1005    &  44    &   12 27 01.33   &    8 20 05.1 \nl
3150    &  21.40    &  1.79   &    952    &  42    &   12 27 05.95   &    8 15 12.1 & &7458    &  20.75    &  1.84   &    807    &  57    &   12 27 12.23   &    8 20 08.6 \nl
3323    &  21.29    &  1.60   &  12321    &  50    &   12 27 26.82   &    8 15 22.8 & &7615    &  21.13    &  1.68   &           &        &   12 27 11.53   &    8 20 20.2 \nl
3412    &  19.83    &  0.72   &   -203    &  38    &   12 27 09.07   &    8 15 30.1 & &7659    &  19.87    &  1.34   &   1571    &  56    &   12 27 10.54   &    8 20 24.0 \nl
3592    &  20.15    &  0.79   &           &        &   12 27 16.91   &    8 15 41.4 & &7731    &  19.55    &  0.82   &    151    &  48    &   12 27 19.33   &    8 20 30.2 \nl
3628    &  21.22    &  1.90   &   1008    &  49    &   12 27 00.59   &    8 15 43.9 & &7784    &  19.20    &  1.52   &    868    &  51    &   12 27 23.22   &    8 20 34.4 \nl
3789    &  19.52    &  1.70   &           &        &   12 27 30.22   &    8 15 53.6 & &7894    &  21.61    &  1.73   &    730    &  81    &   12 27 01.99   &    8 20 37.2 \nl
3808    &  20.35    &  1.83   &    832    &  35    &   12 27 06.60   &    8 15 54.5 & &7938    &  20.92    &  1.44   &   1251    &  50    &   12 27 11.90   &    8 20 47.7 \nl
3865    &  21.60    &  0.80   &           &        &   12 27 02.92   &    8 16 17.2 & &8090    &  20.51    &  1.46   &    903    &  66    &   12 27 13.14   &    8 21 00.9 \nl
3980    &  21.15    &  1.28   &   1112    &  45    &   12 27 03.05   &    8 16 05.4 & &8165    &  20.22    &  1.39   &   1027    &  47    &   12 27 23.57   &    8 21 06.8 \nl
4017    &  20.92    &  1.42   &           &        &   12 27 25.05   &    8 16 07.8 & &8228    &  21.48    &  1.77   &           &        &   12 27 09.19   &    8 21 11.9 \nl
4168    &  20.36    &  1.68   &   1384    &  44    &   12 27 07.64   &    8 16 19.4 & &8353    &  20.03    &  1.98   &    928    &  40    &   12 27 08.71   &    8 21 22.8 \nl
4386    &  19.83    &  1.94   &   1197    &  33    &   12 27 19.16   &    8 16 34.7 & &8384    &  21.39    &  1.41   &    768    &  54    &   12 27 15.30   &    8 21 24.4 \nl
4513    &  20.10    &  1.85   &    908    &  80    &   12 27 09.77   &    8 16 42.7 & &8516    &  21.82    &  2.01   &           &        &   12 27 05.28   &    8 21 36.6 \nl
4542    &  20.62    &  0.68   &           &        &   12 27 22.64   &    8 16 44.5 & &8665    &  19.60    &  0.69   &   -218    &  35    &   12 27 11.86   &    8 21 50.6 \nl
4731    &  19.96    &  1.43   &    698    &  57    &   12 27 09.55   &    8 16 58.1 & &8890    &  20.41    &  1.88   &    870    &  65    &   12 27 15.77   &    8 22 14.0 \nl
4780    &  19.52    &  1.95   &    971    &  45    &   12 27 20.79   &    8 17 00.9 & &8909    &  21.91    &  1.40   &           &        &   12 27 02.41   &    8 22 15.3 \nl
4959    &  21.38    &  1.33   &   1449    &  44    &   12 27 23.05   &    8 17 11.9 & &9009    &  21.28    &  1.99   &           &        &   12 27 12.26   &    8 22 24.1 \nl
5090    &  19.83    &  1.61   &    582    &  46    &   12 27 09.43   &    8 17 20.2 & &9228    &  21.01    &  1.61   &           &        &   12 27 09.08   &    8 22 50.5 \nl
5112    &  20.35    &  0.89   &           &        &   12 27 16.37   &    8 17 21.4 \nl
\enddata
\end{deluxetable}

\clearpage

\begin{deluxetable}{lllrrrrrllr}
\tablenum{3}
\tablecaption{NGC~4472 confirmed
globular clusters.  Successive columns
give ID, T1 magnitude, C-T1 colour, heliocentric velocity, 
velocity error, galactocentric radius, projected radius
along the major axis, [Fe/H], RA(1950),
Dec(1950) and the identification in Table~III of Mould \etal (1990).
\label{tab3}} 
\scriptsize
\startdata
ID  &  T1  &  C-T1  &  V(he)  &  Err  &  r  &  r$_{maj}$  &  
[Fe/H]  &  RA(1950)& Dec(1950) & Other ID  \nl
 & & & & & ($''$)  &  ($''$) \nl 
\tableline
1518  &  19.25  &  1.85  &  1050  &  36  &  248  &  181  &  0.0  &
12 27 07.89 &  8 13 00.5 & 234 \nl 
1712  &  20.36  &  1.34  &  1144  &  40  &  234  &  162  &  -1.2  &  12 27 07.54 &  8 13 18.9 \nl 
2031  &  20.71  &  1.37  &  1352$^{\footnotesize 1}$
 &  65  &  178  &  170  &  -1.2  &  12 27 15.12 &  8 13 46.3 & 142 \nl 
2045  &  20.94  &  1.77  &  857  &  54  &  217  &  129  &  -0.2  &  12 27 06.50 &  8 13 47.7 \nl 
2060  &  20.62  &  1.29  &  1108\tablenotemark{a}
  &  65  &  212  &  132  &  -1.3  &  12 27 07.15 &  8 13 48.3 & 243 \nl 
2140  &  20.45  &  1.80  &  730  &  53  &  196  &  192  &  -0.2  &  12 27 21.64 &  8 13 55.3 \nl 
2163  &  20.15  &  2.01  &  402  &  43  &  208  &  197  &  0.3  &  12 27 23.34 &  8 13 57.6 \nl 
2341  &  20.76  &  1.91  &  1001  &  68  &  267  &  77  &  0.1  &  12 27 00.21 &  8 14 12.9 \nl 
2406  &  20.84  &  2.03  &  1244  &  70  &  149  &  131  &  0.4  &  12 27 13.23 &  8 14 18.4 \nl 
2482  &  21.58  &  2.08  &  767  &  56  &  157  &  111  &  0.5  &  12 27 10.16 &  8 14 24.6 \nl 
2528  &  20.34  &  1.46  &  654\tablenotemark{a}
  &  65  &  139  &  138  &  -1.0  &  12 27 16.79 &  8 14 28.2 & 118 \nl 
2543  &  20.27  &  1.36  &  1199  &  48  &  157  &  153  &  -1.2  &
12 27 20.33 &  8 14 29.3 & 90\nl 
2569  &  20.12  &  1.89  &  1056  &  46  &  144  &  110  &  0.1  &
12 27 11.32 &  8 14 31.7 & 196 \nl 
2634  &  19.70  &  1.56  &  1014  &  57  &  173  &  84  &  -0.7  &
12 27 07.08 &  8 14 37.9 & 245 \nl 
3150  &  21.40  &  1.79  &  952  &  42  &  163  &  46  &  -0.2  &  12 27 05.95 &  8 15 12.1 \nl 
3307  &  20.25  &  1.53  &  1790\tablenotemark{a}
  &  65  &  237  &  148  &  -0.8  &  12 27 29.92 &  8 15 21.3 & 19\nl 
3628  &  21.22  &  1.90  &  1008  &  49  &  222  &  -8  &  0.1  &  12 27 00.59 &  8 15 43.9 \nl 
3808  &  20.35  &  1.83  &  832  &  35  &  134  &  9  &  -0.1  &  12 27 06.60 &  8 15 54.5 \nl 
3980  &  21.15  &  1.28  &  1112  &  45  &  182  &  -17  &  -1.4  &  12 27 03.05 &  8 16 05.4 \nl 
4168  &  20.36  &  1.68  &  1384  &  44  &  112  &  -10  &  -0.4  &  12 27 07.64 &  8 16 19.4 \nl 
4386  &  19.83  &  1.94  &  1197  &  33  &  63  &  29  &  0.2  &  12 27 19.16 &  8 16 34.7 \nl 
4513  &  20.10  &  1.85  &  908  &  80  &  78  &  -22  &  0.0  &  12 27 09.77 &  8 16 42.7 \nl 
4731  &  19.96  &  1.43  &  698  &  57  &  82  &  -38  &  -1.0  &
12 27 09.55 &  8 16 58.1 & 215 \nl 
4780  &  19.52  &  1.95  &  971  &  45  &  88  &  11  &  0.2  &  12
27 20.79 &  8 17 00.9 & 86 \nl 
4959  &  21.38  &  1.33  &  1449  &  44  &  123  &  11  &  -1.3  &  12 27 23.05 &  8 17 11.9 \nl 
5090  &  19.83  &  1.61  &  582  &  46  &  90  &  -59  &  -0.6  &
12 27 09.43 &  8 17 20.2 & 218 \nl 
5323  &  20.33  &  0.82  &  1263\tablenotemark{a}
  &  65  &  53  &  -41  &  -2.5  &  12 27 16.29 &  8 17 33.8 & 130 \nl 
5456  &  19.26  &  1.39  &  737\tablenotemark{a}
  &  65  &  68  &  -66  &  -1.1  &  12 27 12.49 &  8 17 41.4 & 177\nl 
5561  &  20.82  &  1.39  &  903  &  48  &  277  &  -144  &  -1.1  &  12 26 56.84 &  8 17 48.3 \nl 
5629  &  21.09  &  1.36  &  522  &  52  &  82  &  -79  &  -1.2  &  12 27 11.90 &  8 17 52.2 \nl 
6164  &  19.79  &  1.65  &  426  &  30  &  110  &  -110  &  -0.5  &
12 27 12.25 &  8 18 27.1 & 181 \nl 
6231  &  20.77  &  1.81  &  1069\tablenotemark{a}
  &  65  &  110  &  -95  &  -0.1  &  12 27 16.54 &  8 18 32.3 & 127 \nl 
6284  &  19.44  &  1.57  &  569  &  54  &  191  &  -59  &  -0.7  &
12 27 25.39 &  8 18 36.8 & 40\nl 
6294  &  21.02  &  1.64  &  1034  &  84  &  233  &  -170  &  -0.5  &  12 27 01.30 &  8 18 37.7 \nl 
6427  &  21.11  &  1.79  &  1141  &  50  &  128  &  -128  &  -0.2  &  12 27 12.47 &  8 18 47.5 \nl 
6520  &  20.06  &  1.86  &  607  &  57  &  131  &  -114  &  0.0  &
12 27 16.75 &  8 18 53.5 & 121 \nl 
6564  &  20.03  &  1.34  &  1077  &  31  &  147  &  -146  &  -1.2  &
12 27 10.77 &  8 18 57.2 & 206\nl 
6696  &  20.08  &  1.59  &  550  &  52  &  173  &  -107  &  -0.6  &
12 27 21.45 &  8 19 08.5 & 82 \nl 
6721  &  21.09  &  1.73  &  1180  &  45  &  185  &  -174  &  -0.3  &  12 27 07.36 &  8 19 10.5 \nl 
6872  &  20.15  &  1.46  &  870  &  41  &  178  &  -175  &  -1.0  &
12 27 09.21 &  8 19 20.7 & 221 \nl 
6989  &  20.61  &  1.75  &  1071  &  50  &  196  &  -126  &  -0.3  &  12 27 21.98 &  8 19 30.5 \nl 
7197  &  20.94  &  1.50  &  782  &  50  &  208  &  -205  &  -0.9  &  12 27 08.44 &  8 19 48.2 \nl 
7340  &  20.91  &  1.77  &  1308  &  124  &  197  &  -173  &  -0.2  &  12 27 17.39 &  8 19 59.0 \nl 
7399  &  20.35  &  1.40  &  1005  &  44  &  285  &  -253  &  -1.1  &  12 27 01.33 &  8 20 05.1 \nl 
7458  &  20.75  &  1.84  &  807  &  57  &  208  &  -207  &  -0.1  &  12 27 12.23 &  8 20 08.6 \nl 
7659  &  19.87  &  1.34  &  1571  &  56  &  229  &  -229  &  -1.2  &  12 27 10.54 &  8 20 24.0 \nl 
7784  &  19.20  &  1.52  &  868  &  51  &  260  &  -181  &  -0.8  &  12 27 23.22 &  8 20 34.4 \nl 
7889  &  18.84  &  1.58  &  614\tablenotemark{a}
  &  65  &  285  &  -179  &  -0.7  &  12 27 25.49 &  8 20 43.5 & I-20 \nl 
7894  &  21.61  &  1.73  &  730  &  81  &  302  &  -281  &  -0.3  &  12 27 01.99 &  8 20 37.2 \nl 
7938  &  20.92  &  1.44  &  1251  &  50  &  247  &  -245  &  -1.0  &  12 27 11.90 &  8 20 47.7 \nl 
8090  &  20.51  &  1.46  &  903  &  66  &  258  &  -252  &  -0.9  &  12 27 13.14 &  8 21 00.9 \nl 
8165  &  20.22  &  1.39  &  1027  &  47  &  291  &  -210  &  -1.1  &  12 27 23.57 &  8 21 06.8 \nl 
8353  &  20.03  &  1.98  &  928  &  40  &  293  &  -293  &  0.3  &  12 27 08.71 &  8 21 22.8 \nl 
8384  &  21.39  &  1.41  &  768  &  54  &  280  &  -265  &  -1.1  &  12 27 15.30 &  8 21 24.4 \nl 
8890  &  20.41  &  1.88  &  870  &  65  &  329  &  -309  &  0.0  &  12 27 15.77 &  8 22 14.0 \nl 
9991  &  19.41:  &  1.27:  &  1040\tablenotemark{a}
  &  65  &  213  &  175  &  -1.4  &  12 27 26.38 &  8 14 35.4 & 34\nl 
9992  &  19.99:  &  1.47:  &  641\tablenotemark{a}
  &  65  &  34  &  -27  &  -0.9  &  12 27 15.80 &  8 17 16.7 & 135 \nl 
\enddata
\tablenotetext{a}{Velocity from Mould \etal (1990)}
\end{deluxetable}

\clearpage

\begin{deluxetable}{lccccc}
\tablenum{4}
\tablecaption{Results from non-linear fits of
equation 1 to globular cluster samples
taken from Table~\ref{tab3}. Column 4 contains the confidence
level in  $V_{rot}$ obtained by randomizing the position angles of
the observed clusters. Column 5 contains the $2\sigma$ (95\%
confidence) upper limits to the rotation velocity obtained
from Monte-Carlo simulations. The values of $V_{rot}$ in 
parentheses are those obtained when the position angle
of the line of nodes is held fixed at
PA$=162\deg$. \label{tab4}}
\startdata
 &$V_{rot}$ & $\theta_0$ & $V_0$ & Confidence & $V_{rot}^{max}$ (95\%)\nl
 & (kms$^{-1}$)& (deg) & (kms$^{-1}$)& (\%) & (kms$^{-1}$) \nl
\tableline
 All clusters (N=57) & 85 (71) & 125 & 970 & 88 & 150\nl
 Excluding \#19 (N=56) & 67 (45) & 112 & 956 & 78 & 150\nl
\nl
 Blue clusters (N=30) & 159 (142) & 131 & 1038 & 99 & 225\nl
 Excluding \#19 (N=29)& 117 (92) & 122 & 1010 & 96 & 175\nl
\nl
 Red clusters (N=27) & 50 (7) & 83 & 917 & 28 & 100\nl
\enddata
\end{deluxetable}

\clearpage

\clearpage  

\clearpage

\clearpage  

\clearpage

\clearpage  

\clearpage

\clearpage  

\end{document}